\documentclass[10pt,conference]{IEEEtran}
\IEEEoverridecommandlockouts

\usepackage{cite}
\usepackage{amsmath,amssymb,amsfonts}
\usepackage{algorithmic}
\usepackage{graphicx}
\usepackage{textcomp}
\usepackage{xcolor}
\usepackage[hidelinks,bookmarks=true]{hyperref}
\usepackage[numbered]{bookmark}
\usepackage{soul}
\usepackage{booktabs}
\usepackage{multirow}
\usepackage{float}
\usepackage{url}
\usepackage[small,it]{caption}
\usepackage{tcolorbox}
\usepackage{tabularx} 
\usepackage[switch]{lineno}
\usepackage{titlesec}
\usepackage{pgf-pie}
\usepackage{enumitem}
\usepackage{seqsplit}

\setlist[itemize,enumerate]{noitemsep, topsep=0pt, leftmargin=1.0em}
\hypersetup{
	pdfborder          = {0 0 0},
	breaklinks         = true,
	bookmarksnumbered  = true,
	pdfsubject         = {29th IEEE/ACM International Conference on Program Comprehension},
	pdfkeywords        = {Program Comprehension} {Test Method Names} {Rename Refactoring} {Part-of-speech Tags} {Grammar Pattern},
	pdftitle           = {Using Grammar Patterns to Interpret Test Method Name Evolution},
	pdfauthor          = {Anthony Peruma and Emily Hu and Jiajun Chen and Eman Abdullah Alomar and Mohamed Wiem Mkaouer and Christian D. Newman}
}

\newcommand{\RQA}{\textbf{RQ1}: Based on the grammar patterns, how are test methods typically structured, how does this structure evolve, and how does it compare to previously-defined naming patterns?} 
\newcommand{\RQB}{\textbf{RQ2}: How are changes to the grammar pattern related to changes in the semantic meaning of the corresponding method name?} 
\newcommand{\RQC}{\textbf{RQ3}: What are the most common term changes, and what is the relationship between the added term and removed term?}

\begin{document}

\title{\LARGE Using Grammar Patterns to Interpret Test Method Name Evolution 
}

\author{
    \IEEEauthorblockN{Anthony Peruma\IEEEauthorrefmark{1}, Emily Hu\IEEEauthorrefmark{2}, Jiajun Chen\IEEEauthorrefmark{3}, Eman Abdullah Alomar\IEEEauthorrefmark{1}, \\Mohamed Wiem Mkaouer\IEEEauthorrefmark{1}, Christian D. Newman\IEEEauthorrefmark{1}}
    \IEEEauthorblockA{\IEEEauthorrefmark{1}Rochester Institute of Technology, Rochester, NY, USA}
    \IEEEauthorblockA{\IEEEauthorrefmark{2}Tufts University, Medford, MA, USA}
	\IEEEauthorblockA{\IEEEauthorrefmark{3}Stony Brook University, Stony Brook, NY, USA}
    axp6201@rit.edu, emily.hu@tufts.edu, jiajun.chen.2@stonybrook.edu, eman.alomar@mail.rit.edu, \\mwmvse@rit.edu, cnewman@se.rit.edu
}


\maketitle

\begin{abstract}
It is good practice to name test methods such that they are comprehensible to developers; they must be written in such a way that their purpose and functionality are clear to those who will maintain them. Unfortunately, there is little automated support for writing or maintaining the names of test methods. This can lead to inconsistent and low-quality test names and increase the maintenance cost of supporting these methods. Due to this risk, it is essential to help developers in maintaining their test method names over time. In this paper, we use grammar patterns, and how they relate to test method behavior, to understand test naming practices. This data will be used to support an automated tool for maintaining test names.
\end{abstract}


\section{Introduction}
In software, test methods names are constructed to describe both the entity that is being tested as well as actions taken by the test \cite{Zhang2015ASE}. The name of a test is important for the same reason production method names are important; they help developers understand the purpose of the method. Further, these names can be used by automated approaches to analyze/understand test methods and automatically generate code for the test methods. Prior research indicates that test method names have a different structure than production method names \cite{Zhang2015ASE, Wu2020JSS}, but understanding how they are similar or different is still a problem that has not been sufficiently addressed. Prior studies on method naming focus on detecting linguistic anti-patterns \cite{arnaoudova:2015}, method naming bugs \cite{Host2009ECOOP}, and a multitude of naming/renaming practices \cite{Peruma2018IWoR, Peruma2019SCAM, Newman2020JSS, bavota2020, Liu2019ICSE, Arnaoudova2014TSE}, but do not differentiate between production and test method naming structures. For this reason, the concepts discussed in these papers may not fully generalize to test method names. This will hinder our ability to improve and support test method name quality both in the case where they are manually written by developers or automatically generated by tools. It is important to consider the unique structure of test method names to complement and increase the impact of prior work by taking into account the unique structure and purpose of test method names.

We begin addressing this problem by studying the evolution of method name structure and semantics in test suites by, primarily, analyzing the sequence of part-of-speech (POS) tags, called grammar patterns \cite{Newman2020JSS}, associated with the method's name. The purpose of this type of analysis is to understand how the semantics behind the test method name changes and how these semantics correlate with changes to the actual testing behavior, as defined in the code. POS tags are obtained by splitting an identifier name into its constituent words and then annotating the split identifier manually. A grammar pattern provides us with a template-like sequence of POS tags, which are an abstract representation of an identifier's meaning.

One problem with analyzing identifier names is that it is difficult to automatically determine the meaning of words in an identifier and how these words interact with one another. It is even more challenging to take this meaning and use it to understand how it influences, or is influenced by, the behavior of the code. Grammar patterns allow us to perform this analysis more efficiently by broadly categorizing words into their corresponding POS; this allows us to relate words together and, also, we can relate different POS tags with certain types of code behavior \cite{Newman2020JSS}. \textit{The goal of this paper is to understand how test method names are structured, how they evolve in structure and meaning, and how the structure/meaning of these names relate to statically-verifiable code behavior. The data obtained in this study will be used to facilitate test name recommendation and appraisal.} We answer the following research questions:


\textbf{\RQA}
This question helps us understand the common grammar patterns latent in test method names, how they change over time, and how they are related to the code's behavior. We use this data to 1) understand the relationship between code behavior and grammar patterns, 2) compare our findings with prior work that taxonomizes test names at a coarser level; allowing us to determine whether our finer-grain analysis creates more and/or different patterns, and 3) compare against production name grammar patterns to help us pinpoint the differences between test and production naming structures.



\textbf{\RQB}
Grammar patterns provide a way for us to learn the relationship between words, which is more granular than prior approaches, without comparing their specific definitions. This question  explores how changes to the grammar pattern relate to changes in the meaning of a method name using a taxonomy, first defined by Arnaoudova et al. \cite{Arnaoudova2014TSE}.

\textbf{\RQC}
In this research question, we look at the most frequent, concrete changes to words when a test method is renamed without using grammar patterns. The purpose of this is to give us an understanding of how these concrete changes are related to the changes we identify when using grammar patterns and to provide us more information about how these concrete names, and their evolution, relate to code behavior.

The contributions of this study are as follows: (1) a manually annotated dataset of test method grammar patterns (available on our website \cite{ProjectWebiste}), (2) new test name patterns and trends, increasing our understanding of the relationship between test name semantics and implementation, and (3) discussion of how test names evolve structurally and semantically.

\section{Experiment Design}
\label{Section:experiment_design}
\begin{figure*}[t]
 	\centering
 	\includegraphics[trim=0cm 0.2cm 0cm 0cm,clip,scale=0.95]{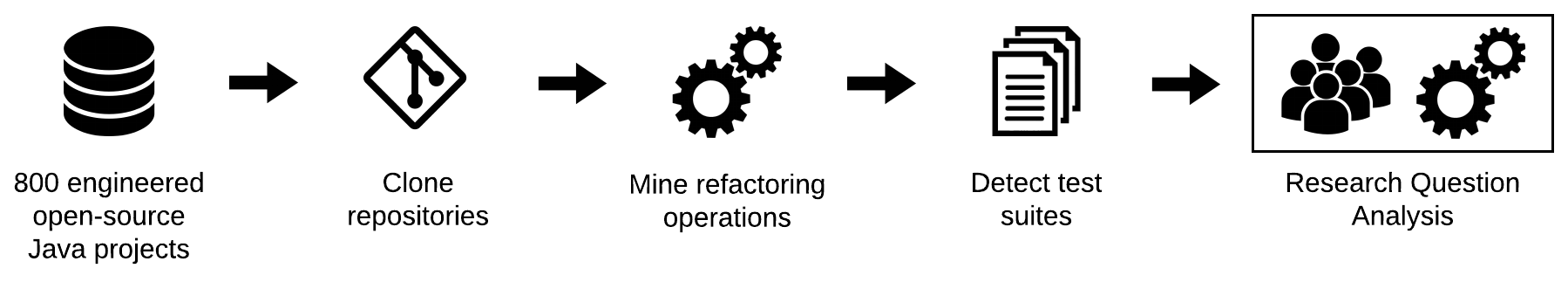}
 	\caption{Overview of our experiment design.}
 	\label{Figure:diagram_experiment}\vspace{-0.5cm}
\end{figure*}

Depicted in Figure \ref{Figure:diagram_experiment} is an overview of our experiment design. We explain, in detail, each activity of our study in the subsequent subsections. Furthermore, the dataset utilized in this study is available on our project website \cite{ProjectWebiste} for extension and replication purposes. 

\vspace{1mm} \noindent \textbf{\textit{Projects:}}
The projects in our study consist of 800 open-source Java projects hosted on GitHub. These projects belong to a curated dataset of engineered software projects, synthesized by the Reaper tool \cite{Munaiah2017ESE}. The projects in this dataset utilize software engineering practices such as documentation, testing, and project management. Not only do we clone each repository, but we also extract commit level metadata by enumerating over the commit log of each project. The metadata we extract includes the timestamp of the commit, the author of the commit, and the files associated with each commit.

\vspace{1mm} \noindent \textbf{\textit{Refactorings:}}
We utilize RefactoringMiner \cite{Tsantalis2018ICSE} for mining the rename refactoring operations from each project in our dataset. RefactoringMiner iterates over the entire commit history of a project in chronological order and compares the changes made to Java source code files in order to detect refactorings. RefactoringMiner is a state-of-the-art tool with a precision of 98\% and a recall of 87\% \cite{Silva2016FSE, Vassallo2019SCP}. Furthermore, we conduct our experiments on the entire commit history of the project (and not on a release-by-release comparison). 

\vspace{1mm} \noindent \textbf{\textit{Test Suites:}}
To identify test suites in the projects, we follow an approach similar to \cite{Peruma2019CASCON}. We first extract all Java source files (i.e., files with extension `.java') that underwent a refactoring. In this study, we focus on projects utilizing the JUnit testing framework \cite{JUnitWebsite}. Next, using JavaParser \cite{JavaParserWebsite}, we parse the Java files by building an abstract syntax tree for each source file. We mark a file as a unit test file if the file contains JUnit import statements (i.e., \texttt{org.junit.*} or \texttt{junit.*}) and a test method. For a file to contain a unit test method, the method should have an annotation called \texttt{@Test} (JUnit 4), or the method name should start with `test' (JUnit 3). In total, we detected 319,108 unit test files, out of which only 12,010 test files had undergone a \textit{Rename Method} refactoring. 

\vspace{1mm} \noindent \textbf{\textit{Research Question Analysis:}}
For our research questions, we make use of the data mined/extracted by the prior activities. The activities involved in answering each research question involve a combination of manual analysis (including data annotation) and quantitative analysis using custom-built tools/scripts. We detail our activities when addressing each research question in Section \ref{Section:experiment_results}, when reporting our results.

\section{Experiment Results}
\label{Section:experiment_results}
In this section, we report on the findings of our experiments.

\subsection*{\RQA}
In this RQ, we examine the POS associated with the old and new names to identify grammar patterns that are specific to test methods. We answer this research question with three sub-RQs. The first sub-RQ looks at the frequently occurring grammar patterns that occur in test method names, while the second sub-RQ compares common grammar prefix patterns reported by prior studies, on test (by Wu and Clause \cite{Wu2020JSS}) and production (by Newman et al. \cite{Newman2020JSS}) method names, with findings from our dataset. Finally, the third sub-RQ examines how the POS changes when the method is renamed. The goal of this RQ, and therefore the sub-RQs, is to identify patterns in the way test method identifiers and their grammar patterns evolve through renames to understand how we can take advantage of this evolution in future research to provide developers with useful feedback about their renaming practices.

\vspace{0.5mm}
\noindent\textbf{\textit{Approach:}} To understand the POS tags that constitute a method name in a unit test file, two authors manually annotated a statistically significant sample of renamed methods. In total, 632 test method rename instances (i.e., the old and new names) were manually analyzed. The analyzed sample represents the original set of 12,010 renamed methods, with a 99\% confidence level and a 5\% confidence interval. Similar to prior research \cite{Newman2020JSS}, our annotation process considered 10 English POS tags-- noun (N), determiner (DT), conjunction (CJ), preposition (P), noun plural (NPL), noun modifier/adjectives (NM), verb (V), verb modifier/adverbs (VM), pronoun (PR), and digit (D). Our annotation process consisted of three stages-- the annotation stage, the review stage, and the discussion stage. In the first stage, each annotator annotated a set of 316 rename pairs of method names. The annotator would first split the old and new method names into their individual terms before annotating each term in the old and new names. Following the annotation process, we conducted a review stage. In this stage, the annotated datasets were exchanged between the annotators for review. If the reviewer did not agree with a specific annotation, the instance was marked for discussion. Finally, in the discussion stage, the annotator and reviewer discussed and looked at resolving conflicts. Instances where there was no consensus were discarded. During the entire process, the annotators had access to the source code and commit diff of the file containing the renamed method to refer. In total, we discarded 17 instances, leaving us with 615 annotated instances for our analysis.

\parindent0pt\subsubsection*{\textbf{RQ 1.1}: What are the most common grammar patterns before and after a rename?}

For this sub-RQ, we look at the frequently occurring grammar patterns for test methods in the annotated dataset (i.e., 615 method renames). These patterns are the complete/full grammar patterns for the names of test methods. Table \ref{Table:CommonOldNewGrammar} shows the top five patterns for old and new names, independent of one another. Below, we elaborate on common grammar patterns.

\begin{table}
\centering
\caption{Top five frequent grammar patterns for old and new names.}
\label{Table:CommonOldNewGrammar}
\begin{tabular}{@{}lrr@{}}
\toprule
\multicolumn{1}{c}{\textbf{Grammar Pattern}} & \multicolumn{1}{c}{\textbf{Count}} & \multicolumn{1}{c}{\textbf{Percentage}} \\ \midrule
\multicolumn{3}{c}{\textit{Old name grammar pattern}} \\
V NM N & 41 & 6.67\% \\
V N & 27 & 4.39\% \\
V NM NM N & 25 & 4.07\% \\
V V NM   N & 22 & 3.58\% \\
V & 20 & 3.25\% \\
\textit{Others} & 480 & 78.05\% \\
\multicolumn{3}{c}{\textit{New name grammar pattern}} \\
V NM N & 44 & 7.15\% \\
V N & 29 & 4.72\% \\
V NM NM N & 29 & 4.72\% \\
V & 17 & 2.76\% \\
NM N & 14 & 2.28\% \\
\textit{Others} & 482 & 78.37\% \\ \bottomrule
\end{tabular}\vspace{-0.9cm}
\end{table}

\textbf{V NM+ N} is also known as a verb phrase pattern; a phrase composed of a verb followed by a noun phrase. Typically, the verb represents the action to be applied to a head-noun that exists within the same phrase; typically the rightmost noun. In test methods, we observe that the term `test' frequently represents the verb. Developers utilize the noun modifier (i.e., adjective) to specify characteristics or context around the entity being tested (i.e., the entity under test). An example of this pattern is the test method \texttt{testStringEncryption} \cite{rq01-a-01}. The term `test' represents the verb or action of the method. The term `Encryption' is the head-noun or the entity under test, while the term `String' represents the noun-modifier; descriptive of the entity under test.

The next two patterns: \textbf{V N} and \textbf{V V NM N} are both derivative verb phrases, where \textbf{V N} is a verb phrase with no adjectives and \textbf{V V NM N} is a verb phrase with an extra verb. Again, the first \textbf{V} is typically the word 'test' or a related term (e.g., can, should; we discuss this later). An example of the \textbf{V N} pattern is \texttt{testParser} \cite{rq01-a-02}, the action is `test', while the term `Parser' represents the object the action is applied to. 

The last pattern is \textbf{V}: This pattern occurs more frequently in test methods than production methods. In production code, these methods have generic names (e.g., `sort') \cite{Newman2020JSS} since they tend to represent generic functionality. However, in test code, the methods falling in this category are part of a test fixture\footnote{A test fixture is utilized by developers to eliminate duplicate code and ensure a fixed environment for the tests.} (i.e., a setup or teardown method). For example, the \texttt{setup} method is utilized by developers to initialize the environment for the test methods in the test suite \cite{rq01-a-03}.

As part of our analysis, we also look for patterns between the terms in the method's name and statements in the method's body. These observations we encounter can be beneficial to static analyzer based code quality tools. These include using the `Assert.fail' method when the method name contains the term `fail' or `failure' (e.g., \texttt{failPrefixMissing} in \cite{rq01-a-04}). Further, the use of the terms `true' and `false' in the method's name is very likely to be associated with using the methods `assertTrue' and `assertFalse' in the method's body, respectively (e.g. \texttt{testUntilTrueDefinitionOnReducedPath} in \cite{rq01-a-05}).

Based on results by Newman et al. \cite{Newman2020JSS} and our study, verb phrases (e.g., \textbf{V NM+ N}) are the most common grammar pattern for method names regardless of whether they are test or production names. Thus, in this sub-RQ, we find no significant difference in the most \textit{frequent} test and production method grammar patterns. However, we also found that approximately 39.29\% of test method grammar patterns are unique (i.e., they only occur once); in contrast to 24.72\% of unique production method grammar patterns \cite{Newman2020JSS}. This difference implies that there may be a more diverse population of patterns in test methods. We address this in the next sub-RQ.


\smallskip
\parindent0pt\subsubsection*{\textbf{RQ 1.2}: How do grammar patterns in test methods compare to defined naming patterns for test and production methods?}

\begin{table*}
\centering
\caption{Occurrence of test naming patterns in test and production code.} 
\label{Table:PrefixPatterns}
\resizebox{\textwidth}{!}{%
\begin{tabular}{@{}llrrrl@{}}
\toprule
\multicolumn{1}{c}{\textbf{\begin{tabular}[c]{@{}c@{}}Wu and Clause's \\ Test Pattern Name\end{tabular}}} & \multicolumn{1}{c}{\textbf{\begin{tabular}[c]{@{}c@{}}Grammar \\ Pattern\end{tabular}}} & \multicolumn{1}{c}{\textbf{\begin{tabular}[c]{@{}c@{}}\# of Old \& New\\ Test Method\\ Instances\end{tabular}}} & \multicolumn{1}{c}{\textbf{\begin{tabular}[c]{@{}c@{}}\% of Test Method\\ Instances Preserved \\ After Rename\end{tabular}}} & \multicolumn{1}{c}{\textbf{\begin{tabular}[c]{@{}c@{}}\# of Production\\ Method \\ Instances\end{tabular}}} & \multicolumn{1}{c}{\textbf{Example}} \\ \midrule
Is and Past Principle Phrase & V V+ & 353 & 62\% & 6 & \begin{tabular}[c]{@{}l@{}}\texttt{testGetActions} \cite{rq01-c-01}\\ `test' and `Get' are verbs\end{tabular} \\ \midrule
Dual Verb Phrase & V V N+ & 52 & 46\% & 0 & \begin{tabular}[c]{@{}l@{}}\texttt{testFindResourceByName} \cite{rq01-c-03}\\ `test' and `Find' are verbs, while `Resource' is a noun\end{tabular} \\ \midrule
\begin{tabular}[c]{@{}l@{}}Verb Phrases \\ With(out) Prepended Test\end{tabular} & V N V+ & 29 & 67\% & 3 & \begin{tabular}[c]{@{}l@{}}\texttt{testFormUploadLargerFile} \cite{rq01-c-04}\\ `test' is a verb, while `Form' is a noun and`Upload' is a verb\end{tabular} \\ \midrule
Divided Duel Verb Phrase & V N V N+ & 2 & 0 & 0 & \begin{tabular}[c]{@{}l@{}}\texttt{testUidFetchBodyPeek} \cite{rq01-c-05}\\ `test' and `Fetch' are verbs, while `Uid' and `Body' are nouns\end{tabular} \\ \midrule
Noun Phrase & N & 5 & 0 & 3 & \begin{tabular}[c]{@{}l@{}}\texttt{main} \cite{rq01-c-06}\\ `main' is a noun\end{tabular} \\ \midrule
Verb With Multiple Nouns Phrase & V N N N & -- & -- & 0 & Not observed in our dataset of annotated test methods\\ \bottomrule
\end{tabular}%
}\vspace{-0.4cm}
\end{table*}

While our findings from the first sub-RQ show us that there are a small number of very frequent grammar pattern which are common to both test and production methods, it also indicates that there may be a difference in the diversity of these patterns. Because we want to understand what grammar patterns tell us about the similarity and differences in test and production methods, we use this sub-RQ to explore common grammar pattern prefixes for test methods; instead of only looking at the full grammar pattern as we did in the prior sub-RQ. By loosening the constraint to allow partial (i.e., prefix) grammar patterns, we aim to understand the diversity of grammar patterns in test methods.


To this end, we compare the catalog of test method name patterns formulated by Wu and Clause \cite{Wu2020JSS} against our annotated dataset, and also examine the occurrence of these patterns in production method names discussed by Newman et al. \cite{Newman2020JSS}. While we compare to Wu and Clause's work, their goals were somewhat different. Their patterns are primarily \textit{prescriptive}; creating templates that developers should use to improve test names. Our work is \textit{descriptive}; attempting to examine the structures latent in test names while not prescribing what developers should use. Even so, the patterns Wu and Clause create are based on testing patterns they observed, and so it is appropriate to relate our patterns to theirs. The difference between our patterns and theirs can be seen in Table \ref{Table:PrefixPatterns}. The leftmost column contains Wu and Clause's pattern names. To its right is another column showing the grammar pattern that corresponds with Wu and Clause's named patterns. Wu and Clause abstract away some detail (i.e., the tail of our grammar patterns) to necessarily and effectively discuss general naming patterns and their semantics. In this paper, we keep these details, which causes several of our patterns to fit into a single one of Wu and Clause's patterns due to being derivative of a high-level pattern they already identified. However, this helps us understand how some of the granular differences that do not appear in Wu and Clause's work affect test name semantics.

In Table \ref{Table:PrefixPatterns}, we also show the frequency of Wu and Clause's patterns in our data, the number of production methods with the corresponding pattern, and the percentage of these patterns, which were conserved after a rename was applied. Where applicable, the `+' symbol, in Table \ref{Table:PrefixPatterns}, indicates that other POS tags precede and/or follow the pattern. One thing to highlight about this table is that we did not find the `Verb With Multiple Phrases' pattern in our dataset, which corresponds to a grammar pattern of \textbf{V N N N}. Part of the reason for this is likely because we used a different tagset than Wu and Clause, who do not appear to use noun modifiers (NM). However, we did not want to assume their tagset and did not find a definition for the tagset they used in their study. Based on our understanding of their patterns, \textbf{V N N N} for them is the same as \textbf{V NM NM N} for our grammar patterns. Also, though we report the frequency at which our patterns match Wu and Clause's, it is important to remember that the tagsets in our studies may not completely match up. This does not matter for our study; we are not trying to determine the legitimacy or frequency of their test patterns. Instead, our work is aiming to find patterns which Wu and Clause may have overlooked and to add further legitimacy to their findings.

The grammar pattern prefixes we find are mostly derivatives of those found by Wu and Clause. However, there are several grammar patterns in our dataset that differ in interesting ways. We discuss these now.


\textbf{V V N P+}: This pattern is similar to Wu and Clause's `Dual Verb Phrase' pattern. The primary difference is the presence of a preposition. For example, in the name \texttt{testReadFileFromClasspath}, `test' and `Read' are verbs, `File' is a noun, and `From' is a preposition \cite{rq01-c-08}. Approximately 43\% of the renames contained the prefix in the old and new names. Some of the common prepositions utilized by developers include `of', `with', and `to'.  Prepositions show a relationship between words, such as when and where things are related to each other. The preposition in this pattern is important because it identifies the relationship between the noun phrases on either side of it. We can use the preposition to assess the quality of a name based on which preposition is used, and whether the behavior of the test supports the use of the provided preposition. For the example above, using static analysis, we can check for the use of a file read operation that leverages the classpath. Normally, this might be difficult, but there is a finite number of prepositions in English (i.e., developers do not create new prepositions on the fly), meaning the behavior they describe is generally well-defined and finite.

\textbf{N V+}: This pattern consists of a noun followed by a verb (e.g., in \texttt{projectClosed}, `project' is a noun, and `Closed' is a verb \cite{rq01-c-09}). Looking at the set of production methods, we observe ten instances of methods with this prefix pattern. From our annotated dataset of test methods, we observe 22 rename instances of this prefix. Additionally, approximately 64\% of the renames contained the prefix in the old and new names.

\textbf{+VM+}: While not strictly a prefix grammar pattern, we include this observation in our findings since 1) verb modifiers have not been discussed at length in prior literature, 2) we found several patterns containing adverbs in our dataset, and 3) it is possible to use some of our observations about naming and implementation practices based on the presence or absence of certain adverbs. We start with an example: in the name \texttt{test\_get\_NotExisting}, `Not' is an adverb \cite{rq01-c-10}). We encounter 86 rename instances containing one or more adverbs in the name. Furthermore, we notice that developers utilize the same adverb from the old name in the new name when performing a rename of the method 78\% of the time. Additionally, the top three terms associated with an adverb are `not' (26 instances), `when' (25 instances), and `exactly' (5 instances). When combined with static code analysis, our observation becomes useful as it helps in appraising the name of an identifier. For instance, when examining the source code, we observe that method names containing the adverb `not' are typically associated with some form of null based checking (e.g. use of `assertNull' in the method \texttt{test\_get\_NotExisting} \cite{rq01-c-10} and the use of `assertNotNull' in the method  \texttt{deleteindexNotExists} \cite{rq01-c-11}). Finally, looking at production methods, we encountered seven instances of methods using this POS within its name.

\textbf{+DT+} : Our rationale for the analysis of determiners is similar to our analysis of the \textbf{+VM+} pattern. Our dataset contains 72 instances that contain determiners in either the old or new name. From this set, there are 42 instances where the developer uses the same determiner in the old and new name (e.g., the term `All' is preserved in the rename \texttt{findAllWithGivenIds} $\rightarrow$ \texttt{findAllWithIds} \cite{rq01-b-05}). Regarding terms, the top three popular determiners are `the', `no', and `all'. In terms of static code analysis, we observe that the term `all' frequently co-occurs with collection-based data types in the method body (e.g., the use of `List$<$Long$>$' in the method \texttt{testExecuteAll} \cite{rq01-b-06}). The static analysis accuracy can be further improved by incorporating Peruma et al. \cite{Peruma2020JSS} findings, specifically the findings on collection-based data types and singular/plural term changes.


Using grammar patterns, we have confirmed the existence of several naming patterns introduced by Wu and Clause. In addition, we identify patterns that were not identified in Wu and Clause's original set of patterns. The patterns we present are not frequent in production method names based on prior research, indicating that they are specific to test method names.

\smallskip
\parindent0pt\subsubsection*{\textbf{RQ 1.3}: What are the most common grammar patterns before and after a rename?}

\begin{table}
\centering
\caption{Top five frequently occurring pairs of complete grammar patterns for renamed unit test methods.}
\label{Table:RenameGrammarPatternTop5}
\begin{tabular}{@{}llrr@{}}
\toprule
\multicolumn{2}{c}{\textbf{Rename Grammar Pattern}} & \multicolumn{1}{c}{\multirow{2}{*}{\textbf{Count}}} & \multicolumn{1}{c}{\multirow{2}{*}{\textbf{Percentage}}} \\
\multicolumn{1}{c}{\textbf{Old Pattern}} & \multicolumn{1}{c}{\textbf{New Pattern}} & \multicolumn{1}{c}{} & \multicolumn{1}{c}{} \\ \midrule
V NM N & V NM N & 14 & 2.28\% \\
V NM NM   N & V NM NM N & 9 & 1.46\% \\
V & V & 7 & 1.14\% \\
V N & V N & 7 & 1.14\% \\
V NM N & V NM NM N & 7 & 1.14\% \\
\multicolumn{2}{c}{\textit{Other Patterns}} & 486 & 92.85\% \\ \bottomrule
\end{tabular}\vspace{-0.5cm}
\end{table}

In this sub-RQ, we examine the evolution of grammar patterns (i.e., the change in the grammar pattern when a method is renamed). In summary, our annotated dataset of 615 rename instances contained 168 (or approximately 27.32\%) rename instances that did not show a change in grammar (i.e., the old and new grammar patterns were the same). Represented in Table \ref{Table:RenameGrammarPatternTop5} are the top five frequently occurring complete grammar pattern pairs. However, looking at the number of instances associated with each pair, we observe a low count (the most being 14 instances). Furthermore, our dataset contained 446 instances of grammar pattern pairs that occurred only once. This phenomenon (i.e., a wide variety of grammar patterns) highlights the diversity of our dataset and, therefore, impacts our analysis of rename pairs. Therefore, for the same purpose as RQ 1.2 we use prefix patterns to perform our analysis. We extracted frequently occurring pairs of rename prefixes for patterns where either the old or new name consists of prefixes of length two, three, four, or five. From this data, we show the top three frequently occurring pairs in Table \ref{Table:Rename-Prefix}; the complete listing is available on our website \cite{ProjectWebiste}.

From these tables, we make a couple of observations. The first is that renames do not typically change the POS tag of a word. Even when a word is changed, it is still the same type (i.e., at the POS level). Further, these renames follow the typical verb phrase method naming grammar pattern \textit{V NM N} $\rightarrow$ \textit{V NM N}. For example, in commit \cite{rq01-a-01}, when renaming the method \texttt{testStringEncryption} $\rightarrow$ \texttt{testStrongEncryption}, the POS is preserved even though terms in the name are changed; the terms `String' and `Strong' are considered as noun modifiers in this instance.

The second observation comes from Table \ref{Table:Rename-Prefix}. As the prefixes in this table increase, the original set of grammar prefixes remain the same. For instance, consider the two prefix pattern \textit{V V} $\rightarrow$ \textit{V V}, when the prefix pattern increases to three, the new pattern still retains the original prefix pattern: \textit{V V NM} $\rightarrow$ \textit{V V NM}. This observation remains consistent as prefixes increase to five prefixes. This shows that grammar pattern prefixes for test method names are consistent across renames. The primary takeaway from this sub-RQ is that grammar patterns are  stable.



\begin{tcolorbox}[top=2mm, bottom=2mm, left=2mm, right=2mm]
\textbf{Summary.}
Using prefix grammar patterns of method renames, we obtained many interesting pattern changes to analyze. We performed this analysis in the context of prior work on test name templates. Our analysis confirms a number of the test name templates and also shows the existence of a few patterns that do not match up to any template provided in prior work. Particularly, patterns that include determiners, prepositions, and adverbs. We find that they have special, oftentimes implementation-oriented meaning in test method names. Finally, in RQ 1.3, we find that grammar pattern prefixes are stable; they do not change very often during rename activities. 

\end{tcolorbox}
\begin{table}
\centering
\caption{Top two frequently occurring pairs of two, three, four and five prefix grammar patterns for renamed unit test methods.}
\label{Table:Rename-Prefix}
\begin{tabular}{@{}llrr@{}}
\toprule
\multicolumn{2}{c}{\textbf{Rename Prefix Grammar Pattern}} & \multicolumn{1}{c}{\multirow{2}{*}{\textbf{Count}}} & \multicolumn{1}{c}{\multirow{2}{*}{\textbf{Percentage}}} \\
\multicolumn{1}{c}{\textbf{Old Pattern}} & \multicolumn{1}{c}{\textbf{New Pattern}} & \multicolumn{1}{c}{} & \multicolumn{1}{c}{} \\ \midrule

\multicolumn{4}{c}{\textit{\textbf{Two Prefix Pattern}}} \\
V V & V V & 142 & 23.51\% \\
V NM & V NM & 103 & 17.05\% \\
V N	& V N & 39 & 6.46\% \\
\multicolumn{2}{r}{\textit{Other Patterns}} & 320 & 52.98\% \\ \midrule

\multicolumn{4}{c}{\textit{\textbf{Three Prefix Pattern}}} \\
V V NM & V V NM & 55 & 9.79\% \\
V NM N & V NM N & 36 & 6.41\% \\
V NM NM	& V NM NM & 29 & 5.16\% \\
\multicolumn{2}{r}{\textit{Other Patterns}} & 442 & 78.65\% \\ \midrule

\multicolumn{4}{c}{\textit{\textbf{Four Prefix Pattern}}} \\
V V NM N & V V NM N & 24 & 4.96\% \\
V NM NM N & V NM NM N & 19 & 3.93\% \\
V V NM NM & V V NM NM & 14 & 2.89\% \\
\multicolumn{2}{r}{\textit{Other Patterns}} & 427 & 88.22\% \\ \midrule

\multicolumn{4}{c}{\textit{\textbf{Five Prefix Pattern}}} \\
V V NM NM N & V V NM NM N & 10 & 2.75\% \\
V V NM N P & V V NM N P & 9 & 2.48\% \\
V V NM NM N	& V NM NM N	& 4	& 1.10\% \\
\multicolumn{2}{c}{\textit{Other Patterns}} & 340 & 93.66\% \\ \bottomrule
\end{tabular}
\end{table} 

\subsection*{\RQB}

\vspace{0.5mm}
\noindent\textbf{\textit{Approach:}} To determine the semantic change a name undergoes during a rename, we utilize a rename taxonomy defined by Arnaoudova et al. \cite{Arnaoudova2014TSE} and utilized in prior identifier rename studies \cite{Peruma2018IWoR, Peruma2019SCAM, Peruma2020JSS} on our annotated dataset. This taxonomy helps us categorize renames into two categories-- renaming form and semantic change. The renaming form looks at the terms added and removed to determine the complexity of the rename-- simple, complex, reordering, and formatting. A rename is simple if only one term is added or removed. A complex change occurs if more than one term is added or removed. Reordering is when two or more terms switch positions. Finally, a formatting change occurs if the developer only makes a change in case or adds/removes a separator or number. In terms of semantic change categories, a rename can either preserve or modify the meaning of the name. A modification to a name can change, narrow, broaden, add or remove the meaning of the name.

\begin{figure}
\centering
\begin{tikzpicture}
\begin{scope}[scale=0.7,xshift=2.2cm]
\pie[] {
41/ Change,
18/ Broaden,
18/ Narrow,
14/ Preserve,
5/ Add,
4/ Remove}
\end{scope}
\end{tikzpicture}
\caption{Proportion of semantic updates to 615 annotated test methods.}
\label{Figure:SemanticPieChart}
\end{figure}
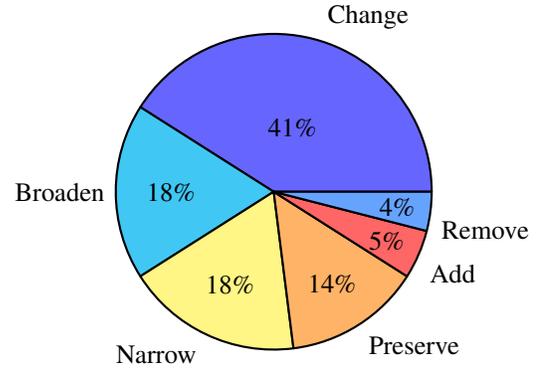

\vspace{0.5mm}From our set of 615 annotated instances, we observe 291 (or approximately 47\%) of the instances had a simple change, while 261 instances (or approximately 42\%) had a complex change. From the semantic category, as depicted in Figure \ref{Figure:SemanticPieChart}, we observe 255 (or approximately 41\%) of instances had a change in meaning, while the narrowing and broadening category each had approximately 18\%. These findings are in contrast to prior research \cite{Peruma2018IWoR,Peruma2019SCAM}, which shows that the majority of renames are of simple form and narrow in meaning. The prior studies utilize datasets comprising of mined rename refactoring operations of test and production source code files in Java projects. Looking at the set of renames categorized under preserve, we observe that the majority of these renames are due to developers either adding or removing numbers or underscore characters to/from the old name or performing a change of case (e.g., \texttt{test\_13} $\rightarrow$ \texttt{test13} \cite{rq03_00}). 

\begin{table}
\centering
\caption{Top five frequently occurring rename semantic updates for pairs of complete grammar patterns.}
\label{Table:SemanticTop5}
\begin{tabular}{@{}lllrr@{}}
\toprule
\multicolumn{2}{c}{\textbf{Rename Grammar Pattern}} & \multicolumn{1}{c}{\multirow{2}{*}{\textbf{Result}}} & \multicolumn{1}{c}{\multirow{2}{*}{\textbf{Count}}} & \multicolumn{1}{c}{\multirow{2}{*}{\textbf{Percentage}}} \\
\multicolumn{1}{c}{\textbf{Old Pattern}} & \multicolumn{1}{c}{\textbf{New Pattern}} & \multicolumn{1}{c}{} & \multicolumn{1}{c}{} & \multicolumn{1}{c}{} \\ \midrule
V NM N & V NM N & Change & 10 & 2\% \\
V D D D & V D D D & Preserve & 6 & 1\% \\
V NM NM N & V NM NM N & Change & 6 & 1\% \\
V N & V NM N & Narrow & 5 & 1\% \\
V NM N & NM N & Broaden & 5 & 1\% \\
\multicolumn{3}{c}{\textit{Other Patterns}} & 583 & 95\% \\
\bottomrule
\end{tabular}\vspace{-0.5cm}
\end{table}

\begin{table}
\centering
\caption{Frequently occurring rename pairs of prefix patterns for different semantic categories.}
\label{Table:Semantic-Prefix}
\begin{tabular}{lllrr}
\hline
\multicolumn{2}{c}{\textbf{Rename Grammar Pattern}} & \multicolumn{1}{c}{\multirow{2}{*}{\textbf{\begin{tabular}[c]{@{}c@{}}Semantic\\ Type\end{tabular}}}} & \multicolumn{1}{c}{\multirow{2}{*}{\textbf{Count}}} & \multicolumn{1}{c}{\multirow{2}{*}{\textbf{Percentage}}} \\
\multicolumn{1}{c}{\textbf{Old Pattern}} & \multicolumn{1}{c}{\textbf{New Pattern}} & \multicolumn{1}{c}{} & \multicolumn{1}{c}{} & \multicolumn{1}{c}{} \\ \hline
\multicolumn{5}{c}{\textit{\textbf{Two Prefix Pattern}}} \\ \hline
V V & V V & Change & 70 & 28.46\% \\
V   NM & V NM & Change & 50 & 20.33\% \\
V N	& V N & Change	& 15	& 6.10\% \\
\multicolumn{3}{c}{\textit{Other Change Patterns}} & 111 & 45.12\% \\ \hline
V V & V V & Preserve & 20 & 22.99\% \\
V   NM & V NM & Preserve & 14 & 16.09\% \\
V N	& V N & Preserve	& 13	& 14.94\% \\
\multicolumn{3}{c}{\textit{Other Preserve Patterns}} & 40 & 45.98\% \\ \hline
V V & V V & Add & 7 & 30.43\% \\
V NM & V NM & Add & 4 & 17.39\% \\
N V	& N V & Add	& 2 & 8.70\% \\
\multicolumn{3}{c}{\textit{Other Add Patterns}} & 10 & 43.48\% \\ \hline
V V & V NM & Remove & 7 & 23.33\% \\
V V & V V & Remove & 4 & 13.33\% \\
V N	& V N	& Remove & 2	& 6.67\% \\
\multicolumn{3}{c}{\textit{Other Remove Patterns}} & 17 & 56.67\% \\ \hline
V V & V NM & Broaden & 19 & 17.27\% \\
V NM & NM N & Broaden & 9 & 8.18\% \\
V NM	& V NM	& Broaden & 9	& 8.18\% \\
\multicolumn{3}{c}{\textit{Other Broaden Patterns}} & 73 & 66.36\% \\ \hline
V V & V V & Narrow & 32 & 29.63\% \\
V NM & V NM & Narrow & 24 & 22.22\% \\
V N	& V NM & Narrow	& 8	& 7.41\% \\
\multicolumn{3}{c}{\textit{Other Narrow Patterns}} & 44 & 40.74\% \\ \hline
\multicolumn{5}{c}{\textit{\textbf{Three Prefix Pattern}}} \\ \hline
V V NM & V V NM & Change & 30 & 13.16\% \\
V NM N & V NM N & Change & 21 & 9.21\% \\
V NM NM	& V NM NM & Change	& 17	& 7.46\% \\
\multicolumn{3}{c}{\textit{Other Change Patterns}} & 160 & 70.18\% \\ \hline
V V NM & V V NM & Preserve & 11 & 14.67\% \\
V NM N & V NM N & Preserve & 7 & 9.33\% \\
V D D	& V D D & Preserve	& 6	& 8.00\% \\
\multicolumn{3}{c}{\textit{Other Preserve Patterns}} & 51 & 68.00\% \\ \hline
V NM N & V NM N & Add & 2 & 8.70\% \\
V NM N & V NM NM & Add & 2 & 8.70\% \\
N NM	& N NM P & Add	& 1	& 4.35\% \\
\multicolumn{3}{c}{\textit{Other Add Patterns}} & 18 & 78.26\% \\ \hline
V V NM & V NM NPL & Remove & 3 & 10.00\% \\
V V NM & V NM N & Remove & 2 & 6.67\% \\
DT NM NM	& NM NM N & Remove	& 1	& 3.33\% \\
\multicolumn{3}{c}{\textit{Other Remove Patterns}} & 24 & 80.00\% \\ \hline
V NM NM & V NM N & Broaden & 8 & 7.69\% \\
V V NM & V NM N & Broaden & 7 & 6.73\% \\
V V NM	& V NM NM & Broaden	& 7	& 6.73\% \\
\multicolumn{3}{c}{\textit{Other Broaden Patterns}} & 82 & 78.85\% \\ \hline
V V NM & V V NM & Narrow & 11 & 10.78\% \\
V NM N & V NM NM & Narrow & 6 & 5.88\% \\
V N	V & NM N & Narrow	& 5	& 4.90\% \\
\multicolumn{3}{c}{\textit{Other Narrow Patterns}} & 80 & 78.43\% \\ \hline
\end{tabular}\vspace{-0.8cm}
\end{table}
Examining the change in meaning instances, 208 (or 33.82\%) of the instances show an unrelated relationship between the old and new names. For example, in renaming \texttt{testLog} $\rightarrow$ \texttt{testEigenSingularValues} \cite{rq02-01}, there is no semantic relationship between the swapped terms. Approximately 7.15\% instances contained more than one type of relationship between the old and new names. For example, in renaming \texttt{testDeserializeExpandCharge} $\rightarrow$ \texttt{testDeserializeWithExpansions} \cite{rq02-03}, we observe an addition and removal of terms as well as a change in plurality. Finally, three (or 0.49\%) instances exhibited an antonym relationship as in case of renaming the method \texttt{shouldAcceptRaxProtocols} $\rightarrow$ \texttt{shouldRejectRaxProtocols} \cite{rq02-02}; here we see the term `Reject' replacing `Accept'.

In Table \ref{Table:SemanticTop5}, we provide the top five rename forms and semantic updates associated with a complete grammar pattern pair. From this table, we observe that the most frequent grammar pair, \textit{V NM N} $\rightarrow$ \textit{V NM N} is mostly associated with a change in meaning. For example, in the rename commit \cite{rq01-a-01}, \texttt{testStringEncryption} $\rightarrow$ \texttt{testStrongEncryption}, a single term is replaced making it a Simple form type change and since there is no semantic relationship between the terms `String' and `Strong' it is categorized as a general change in meaning. However, from this table, we observe a low volume of instances of grammar patterns associated with the semantic categories; this behavior is similar to what is observed in RQ 1.3. Hence, similar to RQ 1.3, going forward, we look at the relationship between prefix grammar patterns and name semantics. 
Presented in Table \ref{Table:Semantic-Prefix}, we provide the top three frequently occurring prefix patterns for each semantic category. The complete listing is available on our website \cite{ProjectWebiste}.

From Table \ref{Table:Semantic-Prefix}, we observe that the rename prefix pattern  \textit{V V} $\rightarrow$ \textit{V V} is associated with all semantic categories. However, it is more prevalent with the change in meaning category. This same prefix pattern is also the most common rename pattern, as reported in RQ 1.3. From the table, we observe that remove and broaden meaning shows a divergence in the prefix pattern; the most frequently occurring prefix pattern for these two categories is \textit{V V} $\rightarrow$ \textit{V NM}. For the broadening pattern, we observed that in the majority of developers tend to remove the term `test' from the old name (e.g., \texttt{testPinnedExternals} $\rightarrow$ \texttt{pinnedExternals} \cite{rq03_01}). 

As the number of prefixes increases, the volume of these instances being associated with a semantic category decreases. Again, similar to RQ 1.3, this phenomenon will help determine the quality of a test method's name either when a developer performs a rename or during static analysis of code. However, as these are prefix patterns, it should be noted that terms associated with the POS tags in the prefix might not always be the terms contributing to the semantic transformation of the method name. Hence, the findings from this RQ should be used in conjunction with findings from other RQs and also prior work, such as Peruma et al. study of identifier renaming using data types and co-occurring refactorings \cite{Peruma2020JSS}.

\begin{tcolorbox}[top=2mm, bottom=2mm, left=2mm, right=2mm]
\textbf{Summary.}
Our analysis of test methods shows that developers frequently change the meaning of a test method's name when performing a rename. This contrasts with prior research, which studied production and test names together, finding that these tend to narrow in meaning. One conclusion we may draw from this is that test methods more frequently change in meaning than the general population of methods. Another potential explanation is that it is more challenging to analyze the relationship between words in test methods. If word relationships in test methods are heavily domain-driven, then some of the underlying technology, such as WordNet, used to analyze these may not work well. More research is needed to conclude which case is valid. However, whichever case we are in, it is clear that the relationship between words in test methods as they evolve is different from the general population of methods. Thus, recommending test name structure or words will potentially require specialized approaches trained specifically on test naming structures.
\end{tcolorbox}

\subsection*{\RQC}

\vspace{0.5mm}
\noindent\textbf{\textit{Approach:}} In this RQ, we examine the frequent terms added to and removed from test methods due to a rename. The experiment in this RQ utilizes the complete dataset of test method names. We first utilize the heuristic splitter algorithm implemented in the Spiral package~\cite{Hucka2018JOSS} to determine the terms that form a name. Next, for each rename instance, we extract only the terms that were added and removed. Finally, for each added and removed pair of terms, we count the number of times the pair exists in the dataset. For example, when \textit{getEmployeeName} is renamed to \textit{testEmployeeLastName}, the added terms are `test' and `Last', while the removed term is `get'. We search our dataset for the occurrence of `test' \& `get' and `Last' \& `get'. Additionally, as part of our qualitative approach, two of the authors manually analyzed a statistically significant sample that comprises of the top 646 frequently occurring pairs. The sample represents a 99\% confidence level and a 5\% confidence interval from our population of 21,615 pairs of added and removed terms. As part of this analysis, the authors annotated the semantic relationships between the added and removed terms. The semantic annotations include: synonyms, antonyms, specializations, and generalizations.

\vspace{0.5mm} 
Analyzing the list of 646 removed-added pairs, we observe instances where the developer either adds or removes numerical digits to or from the replacement term. An in-depth look at these identifiers shows that a vast majority of such names usually do not contain any other terms that describe the behavior of the test method (e.g., \texttt{test15\_6\_5} $\rightarrow$ \texttt{test16\_9\_5} \cite{rq03-000}). Most likely, these are auto-generated tests or tests utilized for debugging purposes. To facilitate the use of English semantic rules to determine the relationship between the term pairs, our analysis of term pairs will be limited to only pairs that do not have numerical digits. Looking at the top five frequently occurring term pairs, we observe developers replace `has' with `contains' (94 instances), `test' with `can' (58 instances), `all of' with `at least' (46 instances), `with' with `when' (40 instances), and `test' with `should' (38 instances). Additionally, we also observe that developers frequently replace the term `test' with a term associated with a Boolean return type (e.g., `can', `is', `should').

Next, we look at the different types of semantic relationships between the removed-added pairs. From our dataset, we observe that 294 of the removed terms were replaced with terms of the same POS. For example, in renaming \texttt{testFilterBaseNice} $\rightarrow$ \texttt{testSelectBaseNice} \cite{rq03-00} the developer replaces the term `Select' with `Filter', both of which are verbs. The majority of replacement terms were added in the same position as the removed term in the name. Looking at the types of semantic relationships in the dataset, we observe 36 pairs of terms as synonyms (e.g., \texttt{boundingCube} $\rightarrow$ \texttt{boundingBox} \cite{rq03-01}) and 22 pairs having an antonym relationship (e.g., \texttt{genericExtension} $\rightarrow$ \texttt{specificExtension} \cite{rq03-02}). We also identified 12 instances each of specialization (e.g.,  \texttt{testPredictions} $\rightarrow$ \texttt{validatePredictions} \cite{rq03-03}) and generalization (e.g., \texttt{listContains} $\rightarrow$ \texttt{collectionContains} \cite{rq03-04}).

Looking at the root (i.e., stem) of the removed-added term pairs, we observe that 70 pair instances have the same stem. For example, in the following rename \texttt{testTwippleUploader} $\rightarrow$ \texttt{testTwippleUpload} \cite{rq03-03}, the terms `Uploader' and `Upload' have the same stem-- `upload'. We also observe 17 instances of tense change (e.g., \texttt{isOrderedFailure} $\rightarrow$ \texttt{isInOrderFailure} \cite{rq03-08}) and nine instances of plurality changes (e.g., \texttt{enqueueJob} $\rightarrow$ \texttt{enqueueJobs} \cite{rq03-07}), and spelling corrections (e.g.,  \texttt{projectVisitorIsInkvoked} $\rightarrow$ \texttt{projectVisitorIsInvoked} \cite{rq03-06}) each.

Our manual analysis shows that determining the relationship between terms is a challenging task due to the diverse ways in which developers rename identifiers. We made the following observations during our qualitative analysis: (1) although proper naming helps understand what the test verifies and how the underlying system behaves, some terms are ambiguous, which makes it challenging to determine the semantic relationship between the pair due to the use of domain terminology (e.g., `LBDevice' is replaced with `Zeus' \cite{rq03-09}), (2) multiple terms can replace a single term and vice versa; this type of change is done due to specialization/generalization of behavior or in situations where the names are synonyms  (e.g., the terms `not started' are replaced by `closed' \cite{rq03-10}), and (3) the terms are unrelated (e.g., `Latency' is replaced with `Metrics' \cite{rq03-11}).

Finally, when examining the code, we observe that specific terms in a method's name can indicate the presence of specific statements in the body of the method. For instance, we observe that the presence of the terms `at least', `all of' or `all' acts as a sign that a method performs tests on collection based objects such as List, Map, or custom collection types. For example, the method \texttt{findAllWithGivenIds} contains a collection object that is subject to a series of tests (i.e., assertion statements). Similarly, the occurrence of the term `exception' indicates that the purpose of such methods is to verify that an exception occurs as part of the execution of the test. In such instances, developers either utilize the `expected' parameter as part of the \texttt{Test} annotation or places an assertion statement in the catch section of the try-catch block that handles the exception that the developer expects to be thrown (e.g., \texttt{\seqsplit{invokingStaticMethodQuietlyShouldWrapIllegalArgumentException}} \cite{rq03-12}). These observations show how static analysis combined with NLP techniques can support the automation of identifier name appraisal algorithms. 


\begin{tcolorbox}[top=2mm, bottom=2mm, left=2mm, right=2mm]
\textbf{Summary.}
When replacing terms in a method's name, developers frequently preserve the overall meaning of the method name by utilizing a synonym of the removed term. In addition, there are some interesting common word and phrase substitutions we observed in this set. Including `has' $\longleftrightarrow$ `contains' and `all of' $\longleftrightarrow$ `at least'. Many of these can be linked with code semantics. For example, `all of' changing to `at least' indicates a shift in testing behavior; instead of testing for the presence of all entities, they are testing for a subset. We manually confirmed that some of this behavior can be directly mapped to code changes, and thus we may be able to provide some naming recommendations in the future based on these trends. In addition, the term `test' is frequently swapped with terms such as `can', `is', and `should'. The relationship between these terms and the term `test' range from synonyms to metonyms.
\end{tcolorbox}


\section{Related Work}
\label{Section:related_work}
We divided our reporting of related work into three areas-- studies that explored the naming of test methods, studies that investigated grammar patterns in identifier names, and studies around the renaming of identifiers in source code.

\subsection{Test Method Names}
Using natural language techniques, Zhang et al. \cite{Zhang2015ASE}, parse test method names in order to generate templates for the test methods automatically. In their approach, the authors utilize the information contained within the name of a test method-- the action phrase and the predicate phrase. To perform the parsing, the authors depend on English grammar constructs. The authors achieve an accuracy of over 80\% for their template generation approach. In a subsequent study \cite{Zhang2016ASE}, the authors present an approach and tool, \textit{NameAssist}, to generate descriptive names for test methods based on the body of the test method. In their approach, the authors analyze the statements within the test method to determine the action, expected outcome, and scenario under test. Based on this static analysis, the authors utilize natural language processing techniques to generate a descriptive name for the test method. 

Daka et al. \cite{Daka2017ISSTA}, propose an approach to generate short descriptive test method names based on  API-level coverage goals; the authors validate their approach by surveying 47 students. Lin et al. \cite{Lin2019SCAM} investigate the quality of identifiers in test suites and perform a comparison against production identifiers. Results from the survey show that identifiers in test suites are of poor quality, with automatically generated test suites demonstrating even more quality concerns. Further, a comparison of rename recommendation tools shows that they perform poorly on test suites. Wu and Clause \cite{Wu2020JSS} provide an approach to identify non-descriptive test method names and provides developers with information for a more descriptive name. Their approach depends on a set of test patterns. From these patterns, their mechanism extracts the action, predicate, and scenario from the current name of the test and body of the test method. By comparing the extracted information, their approach determines if the current test name is descriptive.

\subsection{Identifier Grammar Patterns}
In their study of grammar patterns in identifier names, Newman et al. \cite{Newman2020JSS} observe a set of grammar patterns developers utilize to describe program behavior. Some of their observations include: noun phrases are one of the most common grammar patterns, function identifiers are more likely to be represented by a verb phrase, and collection type frequently utilize a plural head-noun. Additionally, the authors indicate that the current POS taggers are not effective on source code identifiers. 
In an empirical study on 5,000 open-source projects, Zhang et al. \cite{Zhang2020Access} observe that nouns, verbs, and adjectives are three of the most common POS tags utilized by developers in crafting identifier names. The authors utilize Standford Parser to parse the POS tags from an identifier's name automatically. 
Binkley et al. \cite{Binkley2011MSR} investigate the effectiveness of Stanford Log-linear POS Tagger on field names. Through this study, the authors propose four rules, based on POS tags, for improving field names. A study of naming in multiple programming languages by Liblit et al. \cite{Liblit2006APPW} shows how natural language influences the use of words in these languages. H{\o}st and {\O}stvold \cite{Host2009ECOOP} examine unusual method names and propose a set of naming rules to uncover issues in method names. The authors utilize POS tags along with the return type, control flow, and parameters of the method to detect naming violations based on a set of rules.

\subsection{Identifier Renaming}
In a study on identifier renames, Arnaoudova et al. \cite{Arnaoudova2014TSE} propose a semantic taxonomy for classifying identifier renames. 
Additionally, through a developer survey, the authors observe that developers confirm that identifier renaming is a challenge. 

Studies around contextualizing identifier renaming by Peruma et al. \cite{Peruma2018IWoR,Peruma2019SCAM,Peruma2020JSS} show that the majority of identifier renames are performed with the intent of narrowing the meaning of the identifier name. The authors also observe that there is a subset of refactorings that occur before a rename refactoring. Additionally, when looking at data type changes, the authors observe instances where developers change the plurality of a name in response to its type changing to/from a collection. Finally, the authors also discuss challenges around analyzing rename refactorings and commit messages.

Work on identifying rename opportunities by Allamanis et al. \cite{Allamanis2014FSE, Allamanis2015FSE} uses statistical language models to mine natural source code naming conventions. The authors approach looks for potential variable, method, and class renaming opportunities. Research by Liu et al. \cite{Liu2015TSE} look at recommending renames based on the prior rename activities performed by developers on the source code. Additionally, by studying the relationship between argument and parameter names, the authors develop an approach to detect naming anomalies and suggest renames to developers \cite{Liu2016ICSE}. In their study, Jiang et al. \cite{Jiang2019ASE} observe that the effectiveness of \textit{code2vec}, a machine learning-based approach for method name recommendations, fails in a realistic setting. The authors also propose a heuristic-based approach that outperforms \textit{code2vec}.
\section{Threats To Validity}
\label{Section:threats}
The projects in our dataset are limited to open-source Java systems, and the results may not generalize to systems written in other languages. However, these projects are from a set of engineered Java systems \cite{Munaiah2017ESE} and have been utilized in prior refactoring related studies \cite{Peruma2020JSS}. Furthermore, since the test files in our dataset belong to a variety of projects, our analysis is based on test methods implemented by different developers and thereby representative. While there are other tools available to mine refactoring operations, RefactoringMiner outperforms the other tools \cite{Tan2019ASO} and is frequently utilized in refactoring studies \cite{Peruma2019MOBILESoft,AlOmar2020ICSEW,Peruma2020ICSEW,AlOmar2020ESA}. Our analysis of test files is limited to projects utilizing the JUnit testing framework, and hence the results might not generalize to other testing frameworks. However, prior unit testing based research has frequently focused on JUnit, such as in the case of test smells \cite{Garousi2018JSS}. The findings of our research questions are based on a sample set of annotated data. To ensure unbiased representativeness, our sample size is a statistically significant random sample, and our annotation process included a review phase.

\section{Discussion \& Conclusion}
In this paper, we examine how developers craft method names in test suites. We made it our goal to understand how test method names are structured, how they evolve in structure and meaning, and how the structure/meaning of these names relate to statically-verifiable code behavior. Our findings show the effectiveness of using grammar patterns to understand naming practices, and how those practices translate to statically-verifiable code behavior. In this section, we discuss how the findings from our RQ's and observations align with the study's goals, along with the challenges and future research directions.

\subsection{Takeaways}

\subsubsection*{\textbf{Takeaway 1}: Test names have a structure that differs from production names. Some of this structure can be leveraged to provide test-specific recommendations}
From RQ1, we observe that test method names vary in the number of terms making up the name and the POS tags associated with these terms. During our manual annotation, we observe that test method names are highly specific to their intended behavior. This occurrence is not surprising as the purpose of test methods is to test/exercise the atomic units of the production code \cite{Pressman2014Software}; test suites can have more than one test method to test the behavior of a single production method. Hence, developers craft test names to be as descriptive as possible, going as far as even to describe conditions that appear within the method using adverbs (e.g., not) and prepositions (e.g., after) in the test name. The use of these specific adverbs and prepositions is interesting because the appearance of these words correlates with the appearance of specific structures within the code; the words and code structure are semantically related in ways that can be statically detected. This correlation allows for recommendations to be made to developers based on either the presence of certain words in the identifier name or the presence of code structures in the function. The descriptive naming we observe aligns with the action phrase and the predicate phrase naming pattern \cite{Zhang2015ASE}. This phenomenon is highlighted by our findings that show how common prefix grammar patterns for test method names are not common patterns for production methods. 

\smallskip
\subsubsection*{\textbf{Takeaway 2}: Some of the prefixes detected in our dataset indicate the existence of additional test name patterns}
While our findings from RQ1 confirm the grammar patterns identified by Wu and Clause, we also observed test naming patterns for test method names. These grammar patterns are: \textit{V V N P+} and \textit{N V+}. Of these two, \textit{V V N P+} is the only one whose behavior is not sufficiently described by patterns present in prior work, which does not mention prepositions. As stated earlier, prepositions correlate with specific types of test behavior. Thus, we feel that this pattern is both legitimate and new amongst the naming patterns presented in prior research; differentiating this pattern from Wu and Clause's \textit{V V N} is important because it implies different behavior, which a recommendation system should be aware of. In addition, like with prepositions, we find that that the existence of adverbs (VM) and determiners (DT) in test method names correlate with specific code behaviors. This has not been explored in prior work and is novel to our investigation. The evidence we empirically find suggests that the relationship between these POS tags and code behavior can be used to detect and recommend the removal of linguistic anti-patterns.

\smallskip
\subsubsection*{\textbf{Takeaway 3}: Test method name refactorings tend to change the meaning of terms in the name}
By mining rename refactoring operations in projects, we extract renames performed by developers on test methods. This data provides us with the opportunity to study the evolution of method names. From RQ2, we observe that unlike prior research on all types of method names, not focusing exclusively on test suites \cite{Peruma2018IWoR,Peruma2019SCAM}, our study shows that developers frequently change the meaning of test names more often than narrowing the meaning. This indicates one of two situations: either test method names evolve by significantly altering the meaning of words within the name, or it is more challenging to analyze the relationship between words in test methods using approaches such as WordNet than it is to analyze word relationships on other types of methods as was done in prior work \cite{Peruma2018IWoR,Peruma2019SCAM,bavota2020}. Whichever case we are in, it is clear that the relationship between words in test methods as they evolve is different from the general population of methods. Thus, recommending test name structure or words will potentially require specialized approaches trained specifically on test naming structures. More research is needed in this area to understand this phenomenon. 

\smallskip
\subsubsection*{\textbf{Takeaway 4}: There are common words and phrases which are synonymous or metonymous in test method renames}
RQ3 highlights the five most frequently swapped terms in method names during rename operation-- with `has' and `contains' being the most common term pair. These terms/phrases are synonymous, or metonymous, with one another. Further, like the adverb and preposition examples from RQ1, they can be directly linked to specific behaviors that appear within the test code. Thus, it is possible to use these patterns to appraise test names and their contents to provide recommendations. We were only able to identify a handful of these patterns, but they represent a strong start to potential future recommendations and a path toward finding other similar patterns. One clear future step for these terms is to begin taking advantage of the synonym/metonym relationship between them to increase naming consistency in test methods or recommend changes to the test body based on the wording in the test name. This direction is similar to the idea of method name debugging \cite{Host2009ECOOP} and linguistic anti-patterns \cite{arnaoudova:2015}.

\smallskip
\textbf{Summary.} Given the goal of our paper, we have confirmed the usefulness of grammar patterns in understanding test method name structure, how this structure evolves, and how it relates to code behavior. The results from our work help confirm the need for test-specific naming support and provide several recommendations in terms of what kind of support can be readily provided. Specifically, we show that certain types of words are frequently used in the context of statically-verifiable code behaviors. While further research into this phenomenon is required, our work is a strong starting point; providing both the evidence for the existence of these patterns and the means through which these patterns can be further explored, detected, and leveraged for recommendation.

\subsection{Challenges} Key challenges in utilizing standard NLP techniques in studies like ours include parsing of misspellings (e.g., `get' is misspelt as `het' in \texttt{hetInput} \cite{annotate_spelling}), contractions (e.g., expand `dont' to `do not' \cite{annotate_contraction}), domain/technology terms (i.e., terms not part of general purpose ontologies), and preamble terms or terms at the start of the name to indicate a specific action/purpose (e.g., use of  `Ignore' in \texttt{IGNOREtestHttpsCheckOut} to exclude the test \cite{annotate_preamble}). Additionally, analyzing the code and comments surrounding a method will help decide the correct POS tag for terms in its name.

\section{Acknowledgements}
This material is based upon work supported by the National Science Foundation under Grant No. 1850412.

\bibliographystyle{ieeetr}
\bibliography{references}
\end{document}